# The gauge factor increase and the hypothetical emerging of the matter objects on the horizon in the standard model of universe

**Vladimír Skalský**

Faculty of Materials Science and Technology of the Slovak Technical University, 917 24 Trnava, Slovakia,
skalsky@mtf.stuba.sk

**Abstract.** In the standard model of universe the increase in mass of our observed expansive and isotropic relativistic Universe is explained by the hypothetical assumption of matter objects emerging on the horizon (of the most remote visibility). However, the mathematical-physical analysis of the increase of Universe gauge factor shows that this hypothetical assumption is non-compatible with the variants of the standard model of universe by which – according to the standard model of universe – can be described the expansive evolution of the Universe.

## 1. Introduction

The mathematical-physical basis of the present relativistic cosmology represents *the Friedmann general equations of homogeneous and isotropic universe dynamics* (Friedmann 1922, 1924). Using *the Robertson-Walker metrics of homogeneous and isotropic universe* (Robertson 1935, 1936a, b; Walker 1936) they can be expressed in the form:

$$\dot{a}^2 = \frac{8\pi G a^2 \rho}{3} - kc^2 + \frac{\Lambda a^2 c^2}{3}, \tag{1a}$$

$$2a\ddot{a} + \dot{a}^2 = -\frac{8\pi G a^2 p}{c^2} - kc^2 + \Lambda a^2 c^2, \tag{1b}$$

where $a$ is the gauge factor (the overall scale of the universe), $\rho$ is the mass density, $k$ is the curvature index, $\Lambda$ is the cosmological member and $p$ is the pressure.

According to *the Planck quantum hypothesis* (Planck 1899), it has a sense to think about the physical parameters of the observed expansive and isotropic relativistic Universe from the moment when its dimensions reach the values which corresponds to

*the Planck length* $$l_P = ct_P = \sqrt{\frac{\hbar G}{c^3}} = 1.616\,05 \times 10^{-35} \text{ m}, \tag{2}$$

i.e. at

*the Planck time* $$t_P = \frac{l_P}{c} = \sqrt{\frac{\hbar G}{c^5}} = 5.390\,56 \times 10^{-44} \text{ s}. \tag{3}$$

According to *the standard model of universe*, the present properties of the Universe are dependent on its beginning assumptions. The dimensions of the Universe from the beginning of its expansive evolution have been increased approximately by 60 ranges (Skalský 1997, 1999), therefore – according to the standard model of universe – at the Planck time $t_P$ (3) the Universe had to have *the beginning mass density* $\rho_{beg} = (1 \pm 10^{-59})\rho_c$, where $\rho_c$ is the critical mass density. Because, if $\rho_{beg}$ had been larger, its expansive evolution would have been replaced by the contraction. If $\rho_{beg}$ had been smaller, no hierarchic gravitationally-bounded rotational systems (HGRS) in the Universe would have arise, i.e. no life would have arise in the Universe.

It also means that at *the cosmological time of the beginning of radiation (photon) era* $t_{rad} \sim 1$ s the Universe had to have *the mass density at the beginning of radiation era* $\rho_{rad} = (1 \pm 10^{-16})\rho_c$, and at *the cosmological time of the beginning of matter era* $t_{mat} \sim 3 \times 10^5$ yr the Universe had to have *the mass density at the beginning of matter era* $\rho_{mat} = (1 \pm 10^{-3})\rho_c$.

Therefore – according to the standard model of universe – the value of curvature index $k$ up to the cosmological time $t \approx 10^9$ yr we can regard as zero, and the present value of curvature index $k$ of our Universe we do not know.

The dependence of the present properties of Universe from its beginning assumptions in the standard model of universe gives restrictions not only on the value of mass density $\rho$, but on all possible factors which would have influence on the evolution of properties of the expansive homogeneous and isotropic relativistic Universe, i.e. gives restrictions on the value of hypothetical cosmological member $\Lambda$, too. In the present cosmological literature is mostly given that in the observed Universe proportion so-called *the total cosmological constant* on the expansive evolution of Universe cannot be bigger than $10^{-120}$ (Abbot 1985). Therefore, in the variants of the standard universe model in most cases it is shown the value of cosmological member $\Lambda = 0$.

According to the standard model of universe from the pre-inflationary period of the cosmology, the model properties of Universe at the cosmological time $t = t_P$ (3) up to the cosmological time $t \sim 1$ s were determined by the Friedmann equations (1a) and (1b) with the value of curvature index $k = 0$ and the value of cosmological member $\Lambda = 0$ and

*the boundary hard state equation* $\qquad\qquad p = \varepsilon$ . $\qquad\qquad$ (4)

The standard model of universe complemented by *the inflationary evolution phase* from the pre-inflationary variant of the standard model of universe is different only by this that in the beginning era in its expansive evolution proceeds *the inflation*. In all variants of the inflationary universe models past finish the inflationary evolution phase the universe evolutes, according to the variant of the standard model of universe from the pre-inflationary period of the cosmology. According to *the Linde model of chaotic inflationary Universe* (i.e. in the only model of the inflationary models of universe which is considered as viable), the inflationary evolution of Universe ends at the cosmological time $t \sim 10^{-37}$ s (Linde 1990).

According to the standard model of universe, at *the radiation (photon) era*, i.e. at the cosmological time $t \sim 1$ s up to the cosmological time $t \sim 3 \times 10^5$ yr, the Universe can be described by the Friedmann equations (1a) and (1b) with $k = 0$ and $\Lambda = 0$ and

*the ultra-relativistic state equation* $\qquad\qquad p = \frac{1}{3}\varepsilon$ . $\qquad\qquad$ (5)

According to the standard model of universe, at the beginning period of *the matter era*, i.e. at the cosmological time $t \sim 3 \times 10^5$ yr up to the cosmological time $t \approx 10^9$ yr, the Universe can be described by the Friedmann equations (1a) and (1b) with $k = 0$ and $\Lambda = 0$ and

*the dust state equation* $\qquad\qquad p = 0$ , $\qquad\qquad$ (6)

and at the present time, i.e. at the cosmological time $t \sim 1{,}5 \times 10^{10}$ yr, the Universe can be described by the Friedmann equations (1a) and (1b) with unknown value of curvature index $k$, the value of cosmological member $\Lambda = 0$ and the dust state equation (6).

## 2. The velocity of increase of Universe gauge factor, according to the standard model of universe

For the gauge factor $a$, the chosen scale $a_0$ and the cosmological time $t$ of the variant of the standard model of universe, determined by the Friedmann equations (1a) and (1b) with $k = 0$ and $\Lambda = 0$ and the boundary hard state equation (4), is valid the relation (Monin *et al.* 1989):

$$a = \sqrt[3]{3ca_0^2 t} \ . \qquad (7)$$

In all Friedmannian models of expansive universe, the values of gauge factor $a$ at the cosmological time $t$ we can interpret by the relation:

$$a = vt , \qquad (8)$$

where $v$ is the velocity of gauge factor increase.

In the relation (7) – and in the relations (17), (20), (24), (29), (33), (37), (41) and (45) – the gauge factor $a$ at given cosmological time $t$ is dependent on the value of chosen scale $a_0$. Therefore, on the value of chosen scale $a_0$ is dependent also the velocity of gauge factor increase $v$, determined by the relation (8).

In *the very early Universe* the matter-space-time was quantized, therefore, neither the chosen scale $a_0$ nor the cosmological time $t$ at the beginning era of Universe evolution we can not fixed arbitrarily but only as the multiple of the Planck length $l_P$ (2) and the Planck time $t_P$ (3). However, any restrictions, which inhibit the quantization of the space and the time extend to the whole expansive evolution of the Universe, do not exist.

In each from the variants of the standard model of universe the values of gauge factor increase $v$, determined by the relation (8), are equally with all values of the gauge factor $a$, determined by using the values of the chosen scale $a_0 = x l_P$ at the cosmological time $t = t_0 = x t_P$, where $x$ is an arbitrary positive number.

The relation (7) with the values of chosen scale $a_0 = l_P, 2l_P, 3l_P, \ldots 2.998 \times 10^{-29}$ m $= 1.855 \times 10^6 l_P, \ldots$ $2.998 \times 10^8$ m $= 1.855 \times 10^{43} l_P$ at the corresponding cosmological times $t_0 = t_P, 2t_P, 3t_P, \ldots 10^{-37}$ s $= 1.855 \times 10^6 t_P, \ldots 1$s $= 1.855 \times 10^{43} t_P$ give the values of gauge factor

$$a = \sqrt[3]{3cl_P^2 t_P} = 2.331 \times 10^{-35} \text{ m} = 144.225\% \, l_P , \qquad (9)$$

$$a = \sqrt[3]{3c(2l_P)^2 2t_P} = 4.661 \times 10^{-35} \text{ m} = 288.450\% \, l_P , \qquad (10)$$

$$a = \sqrt[3]{3c(3l_P)^2 3t_P} = 6.992 \times 10^{-35} \text{ m} = 432.675\% \, l_P \,, \tag{11}$$

...

$$a = \sqrt[3]{3c(1.855 \times 10^6 l_P)^2 (1.855 \times 10^6 \, t_P)} = 4.323 \times 10^{-29} \text{ m} = 2.676 \times 10^6 \, l_P \,, \tag{12}$$

...

$$a = \sqrt[3]{3c(1.855 \times 10^{43} l_P)^2 (1.855 \times 10^{43} \, t_P)} = 4.324 \times 10^8 \text{ m} = 2.676 \times 10^{43} \, l_P \,. \tag{13}$$

The values in the relations (9), (10), (11), ... (12), ... and (13) we can express in the generalised form, too:

$$a = \sqrt[3]{3c a_0^2 t_0} = 144.225\% \, a_0 \,. \tag{14}$$

From the relations (8), (9), (10), (11), ... (12), ... (13) it results the velocity

$$v = \frac{a\,(9)}{t_P} = \frac{a\,(10)}{2t_P} = \frac{a\,(11)}{3t_P} = ... = \frac{a\,(12)}{10^{-37}\text{s}} = ... = \frac{a\,(13)}{1\,\text{s}} = 4.324 \times 10^8 \text{ m s}^{-1} = 144.225\% \, c \,, \tag{15}$$

which – according to the standard model of universe – expand the Universe in the distance of gauge factor $a$ at the beginning era of its expansive evolution.

If we substitute the relations for the gauge factors $a$ (9), $a$ (10), $a$ (11), ... $a$ (12), ... $a$ (13) and corresponding to them the cosmological times $t_P$, $2t_P$, $3t_P$, ... $1.855 \times 10^6 \, t_P$, ... $1.855 \times 10^{43} \, t_P$ in the relation (15) by the relation for the generalised gauge factor $a$ (14) and the generalised cosmological time $t_0$ we receive the relation (15) expressed in the generalised form:

$$v = \frac{a\,(14)}{t_0} = 4.324 \times 10^8 \text{ m s}^{-1} = 144.225\% \, c \,. \tag{15}$$

However, for the velocity of gauge factor increase $v$ during the whole expansive evolution of Universe is in the standard model of universe shown the value (Linde 1990):

$$v \sim c \tag{16}$$

and is justify by the hypothetical emerging of matter objects on the horizon (of the most remote visibility), i.e. at *the boundary (maximum) velocity of signal propagation*. Therefore, the velocity of increase of gauge factor $a$ in the beginning era of expansive evolution of Universe by the supraluminal velocity $v$ (15) by the hypothetical emerging of matter objects on the horizon (of the most remote visibility), firmly is impossible to explain. Moreover, the values of gauge factor $a$ in the relations (9), (10), (11), (12) and (13), or (14), not justify the assumption of the quantization of the space and the time in the very early Universe, too (Skalský 2000).

According to the standard model of universe, the very early Universe could be determined only by the boundary hard state equation (4). However, if we assume that the beginning period of the expansive evolution of the Universe was not determined by the boundary hard state equation (4) but by the ultra-relativistic state equation (5) then for the determination of the gauge factor $a$ using the chosen scale $a_0$ and the corresponding cosmological time $t$ we must use the relation (Monin *et al.* 1989):

$$a = \sqrt{2c a_0 t} \,, \tag{17}$$

which is valid in the variant of the standard model of universe, determined by the Friedmann equations (1a) and (1b) with $k = 0$ and $\Lambda = 0$ and the ultra-relativistic state equation (5).

The relation (17) with the values of chosen scale $a_0 = l_P$, ... $2.838 \times 10^{21}$ m $= 1.756 \times 10^{56} \, l_P$ and corresponding of the cosmological time $t_0 = t_P$, ... $3 \times 10^5$ yr $= 9.467 \times 10^{12}$ s $= 1.756 \times 10^{56} \, l_P$ gives the values of gauge factor

$$a = \sqrt{2c a_0 t_0} = 141.421\% \, a_0 \tag{18}$$

and – according to the relation (8) – it grows at the velocity

$$v = \frac{a\,(18)}{t_0} = 4.240 \times 10^8 \text{ m s}^{-1} = 141.421\% \, c \,. \tag{19}$$

It means that nor the standard model of universe, determined by the Friedmann equations (1a) and (1b) with $k = 0$ and $\Lambda = 0$ and the ultra-relativistic state equation (5), do not satisfy the hypothetical assumption about emerging of matter objects on the horizon (of the most remote visibility) and do not consider nor the quantization of the space and the time in the very early Universe, too (Skalský 2000).

If we assume that the Universe during the whole expansive evolution is determined by the dust state equation (6) then for the determination of the gauge factor $a$ using the chosen scale $a_0$ and the corresponding cosmological time $t$ we must use the relation (Monin *et al.* 1989):

$$a = \sqrt[3]{\frac{9a_0 c^2 t^2}{4}}, \tag{20}$$

which is valid in the variant of the standard model of universe, determined by the Friedmann equations (1a) and (1b) with $k = 0$ and $\Lambda = 0$ and the dust state equation (6).

The relation (20) with the values of the chosen scale $a_0 = l_P, \ldots 1.429 \times 10^{26}$ m $= 8.781 \times 10^{60} \, l_P$ and the corresponding cosmological time $t_0 = t_P, \ldots 1.5 \times 10^{10}$ yr $= 4.734 \times 10^{17}$ s $= 8.781 \times 10^{60} \, t_P$ give the values of gauge factor

$$a = \sqrt[3]{\frac{9a_0 c^2 t_0^2}{4}} = 131.037\% \, a_0 \tag{21}$$

and – according to the relation (8) – it grows at the velocity

$$v = \frac{a\,(21)}{t_0} = 3.928 \times 10^8 \text{ m s}^{-1} = 131.037\% \, c. \tag{22}$$

It means that nor the standard model of universe, determined by the Friedmann equations (1a) and (1b) with $k = 0$ and $\Lambda = 0$ and the dust state equation (6) do not satisfy the hypothetical assumption about emerging of matter objects on the horizon (of the most remote visibility) and do not satisfy nor the quantization of the space and the time in the very early Universe, too (Skalský 2000).

During the expansive evolution of Universe – according to the standard model universe – approximately at the cosmological time $t > 10^9$ yr could apply two more possibilities: the variant of the standard model of universe, determined by the equations (1a) and (1b) with $k = -1$ and $\Lambda = 0$ and the dust state equation (6) and the variant of the standard model of universe, determined by the equations (1a) and (1b) with $k = +1$ and $\Lambda = 0$ and the dust state equation (6).

In the variant of the standard model of universe, determined by the Friedmann equations (1a) and (1b) with $k = -1$ and $\Lambda = 0$ and the dust state equation (6), for the cosmological time $t$, the chosen scale $a_0$ and the dimensionless conform time $\eta$ is valid the relation (Monin *et al.* 1989):

$$t = \frac{a_0}{2c}(\sinh\eta - \eta) \tag{23}$$

and for the gauge factor $a$, the chosen scale $a_0$ and the dimensionless conform time $\eta$ is valid the relation (Monin *et al.* 1989):

$$a = \frac{a_0}{2}(\cosh\eta - 1), \tag{24}$$

where the dimensionless conform time $\eta$ is defined by the relation:

$$\eta = \pm c \int \frac{dt}{a(t)}. \tag{25}$$

The relation (24) using the relation (23) with the values of chosen scale $a_0 = l_P, \ldots 1.429 \times 10^{26}$ m $= 8.781 \times 10^{60} \, l_P$ and the corresponding cosmological time $t_0 = t_P, \ldots 1.5 \times 10^{10}$ yr $= 4.734 \times 10^{17}$ s $= 8.781 \times 10^{60} \, t_P$ give the values of gauge factor

$$a = \frac{a_0}{2}(\cosh\eta_0 - 1) = 162.198\% \, a_0, \tag{26}$$

and – according to the relation (8) – it grows at the velocity

$$v = \frac{a\,(26)}{t_0} = 4.863 \times 10^8 \text{ m s}^{-1} = 162.198\% \, c. \tag{27}$$

It means that nor the standard model of universe, determined by the Friedmann equations (1a) and (1b) with $k = -1$ and $\Lambda = 0$ and the dust state equation (6), do not satisfy the hypothetical assumption about emerging of matter objects on the horizon (of the most remote visibility).

In the variant of the standard model of universe, determined by the Friedmann equations (1a) and (1b) with $k = +1$ and $\Lambda = 0$ and the dust state equation (6), for the cosmological time $t$, the chosen scale $a_0$ and the dimensionless conform time $\eta$ is valid the relation (Monin *et al.* 1989):

$$t = \frac{a_0}{2c}(\eta - \sin\eta) \tag{28}$$

and for the gauge factor $a$, the chosen scale $a_0$ and the dimensionless conform time $\eta$ is valid the relation (Monin *et al.* 1989):

$$a = \frac{a_0}{2}(1 - \cos\eta). \tag{29}$$

The relation (29) using the relation (28) with the values of chosen scale $a_0 = l_P$, ... $1.429 \times 10^{26}$ m = $8.781 \times 10^{60}\, l_P$ and the corresponding cosmological time $t_0 = t_P$, ... $1.5 \times 10^{10}$ yr = $4.734 \times 10^{17}$ s = $8.781 \times 10^{60}\, t_P$ give the values of gauge factor

$$a = \frac{a_0}{2}(1 - \cos\eta_0) = 91.619\%\, a_0, \tag{30}$$

and – according to the relation (8) – it grows at the velocity

$$v = \frac{a\,(30)}{t_0} = 2.747 \times 10^8\ \text{m s}^{-1} = 91.619\%\, c. \tag{31}$$

It means that nor the standard model of universe, determined by the Friedmann equations (1a) and (1b) with $k = +1$ and $\Lambda = 0$ and the dust state equation (6), do not satisfy the hypothetical assumption about emerging of matter objects on the horizon (of the most remote visibility), because the velocity of gauge factor increase $a$ in it is very small.

### 3. The velocity of gauge factor increase in other variants of the standard model of universe

For completeness we show the relations which determined the velocity of gauge factor increase $a$ in others four variants of the standard model of universe, though by they – according to the standard model of universe – is not possible described nor one from the evolutionary eras of the expansive Universe.

In the variant of the standard model of universe, determined by the Friedmann equations (1a) and (1b) with $k = -1$ and $\Lambda = 0$ and the hard state equation (4), for the cosmological time $t$, the chosen scale $a_0$ and the dimensionless conform time $\eta$ is valid the relation (Monin *et al.* 1989):

$$t = \frac{a_0}{c}\int_0^\eta \sqrt{\sinh 2\xi}\, d\xi \tag{32}$$

and for the gauge factor $a$, the chosen scale $a_0$ and the dimensionless conform time $\eta$ is valid the relation (Monin *et al.* 1989):

$$a = a_0 \sqrt{\sinh 2\eta}. \tag{33}$$

The relation (33) using the relation (32) with the values of chosen scale $a_0 = l_P$, ... $1.429 \times 10^{26}$ m = $8.781 \times 10^{60}\, l_P$ and the corresponding cosmological time $t_0 = t_P$, ... $1.5 \times 10^{10}$ yr = $4.734 \times 10^{17}$ s = $8.781 \times 10^{60}\, t_P$ give the values of the gauge factor

$$a = a_0 \sqrt{\sinh 2\eta_0} = 182.035\%\, a_0 \tag{34}$$

and – according to the relation (8) – it grows at the velocity

$$v = \frac{a\,(34)}{t_0} = 5.457 \times 10^8\ \text{m s}^{-1} = 182.035\%\, c. \tag{35}$$

In the variant of the standard model of the universe, determined by the Friedmann equations (1a) and (1b) with $k = +1$ and $\Lambda = 0$ and the hard state equation (4), for the cosmological time $t$, the chosen scale $a_0$ and the dimensionless conform time $\eta$ is valid the relation (Monin *et al.* 1989):

$$t = \frac{a_0}{c}\int_0^\eta \sqrt{\sin 2\xi}\, d\xi \tag{36}$$

and for the gauge factor $a$, the chosen scale $a_0$ and the dimensionless conform time $\eta$ is valid the relation (Monin *et al.* 1989):

$$a = a_0 \sqrt{\sin 2\eta}. \tag{37}$$

The relation (37) using the relation (36) with the values of chosen scale $a_0 = l_P$, ... $1.429 \times 10^{26}$ m = $8.781 \times 10^{60} l_P$ and the corresponding cosmological time $t_0 = t_P$, ... $1.5 \times 10^{10}$ yr = $4.734 \times 10^{17}$ s = $8.781 \times 10^{60} t_P$ give the values of gauge factor

$$a = a_0 \sqrt{\sin 2\eta_0} = 81.003\% \, a_0 \tag{38}$$

and – according to the relation (8) – it grows at the velocity

$$v = \frac{a\,(38)}{t_0} = 2.428 \times 10^8 \text{ m s}^{-1} = 81.003\% \, c . \tag{39}$$

In the variant of the standard model of universe, determined by the Friedmann equations (1a) and (1b) with $k = -1$ and $\Lambda = 0$ and the ultra-relativistic state equation (5), for the cosmological time $t$, the chosen scale $a_0$ and the dimensionless conform time $\eta$ is valid the relation (Monin *et al.* 1989):

$$t = \frac{a_0}{c}(\cosh\eta - 1) \tag{40}$$

and for the gauge factor $a$, the chosen scale $a_0$ and the dimensionless conform time $\eta$ is valid the relation (Monin *et al.* 1989):

$$a = a_0 \sinh\eta . \tag{41}$$

The relation (41) using the relation (40) with the values of chosen scale $a_0 = l_P$, ... $1.429 \times 10^{26}$ m = $8.781 \times 10^{60} l_P$ and the corresponding cosmological time $t_0 = t_P$, ... $1.5 \times 10^{10}$ yr = $4.734 \times 10^{17}$ s = $8.781 \times 10^{60} t_P$ give the values of gauge factor

$$a = a_0 \sinh\eta_0 = 173.205\% \, a_0 \tag{42}$$

and – according to the relation (8) – it grows at the velocity

$$v = \frac{a\,(42)}{t_0} = 5.193 \times 10^8 \text{ m s}^{-1} = 173.205\% \, c . \tag{43}$$

In the variant of the standard model of universe, determined by the Friedmann equations (1a) and (1b) with $k = +1$ and $\Lambda = 0$ and the ultra-relativistic state equation (5), for the cosmological time $t$, the chosen scale $a_0$ and the dimensionless conform time $\eta$ is valid the relation (Monin *et al.* 1989):

$$t = \frac{a_0}{c}(1 - \cos\eta) \tag{44}$$

and for the gauge factor $a$, the chosen scale $a_0$ and the dimensionless conform time $\eta$ is valid the relation (Monin *et al.* 1989):

$$a = a_0 \sin\eta . \tag{45}$$

The relation (45) using the relation (44) with the values of chosen scale $a_0 = l_P$, ... $1.429 \times 10^{26}$ m = $8.781 \times 10^{60} l_P$ and the corresponding cosmological time $t_0 = t_P$, ... $1.5 \times 10^{10}$ yr = $4.734 \times 10^{17}$ s = $8.781 \times 10^{60} t_P$ give the values of gauge factor

$$a = a_0 \sin\eta_0 = 100\% \, a_0 \tag{46}$$

and – according to the relation (8) – it grows at the velocity

$$v = \frac{a\,(46)}{t_0} = 2.997\,924\,58 \times 10^8 \text{ m s}^{-1} = 100\% \, c . \tag{47}$$

It means that only this variant of the standard model of universe satisfy the hypothetical assumption about emerging of matter objects on the horizon (of the most remote visibility) and satisfy the quantization of the space and the time in the very early Universe, too.

## 4. Conclusions

From nine variants of the standard model of universe only in the Friedmannian model of universe, determined by the Friedmann equations (1a) and (1b) with $k = +1$, $\Lambda = 0$ and the state equation (5), the gauge factor $a$ increases at the velocity $v = c$ which satisfies the assumption of the quantization of the space and the time in the very early Universe and the assumption which results from the hypothetical assumption of the standard model of universe about emerging the matter objects on the horizon (of the most remote visibility) which is determined by the relation (16). However, this model has only formal meaning because detailer analysis shows that this model is determined by the Friedmann equations (1a) and (1b) with $k = +1$, $\Lambda = 0$ and the state equation

$$p = \frac{1}{3}\varepsilon = 0. \tag{44}$$